\documentclass[runningheads]{llncs}
\usepackage[T1]{fontenc}
\usepackage{amsmath, amsfonts, graphicx, booktabs, tabularx, multirow, siunitx, xcolor, soul, bm, enumitem, tikz, mathtools, subcaption}
\usepackage{hyperref}
\usepackage{enumitem}
\usepackage{color}

\graphicspath{{./figures/}}

\begin{document}
\title{HartleyMHA: Self-Attention in Frequency Domain for Resolution-Robust and Parameter-Efficient 3D Image Segmentation\thanks{This paper was accepted by the International Conference on Medical Image Computing and Computer-Assisted Intervention (MICCAI 2023). The final publication is available at Springer via \url{https://doi.org/10.1007/978-3-031-43901-8_35}.}}
\titlerunning{HartleyMHA: Self-Attention in Frequency Domain}
%
\author{Ken C. L. Wong, Hongzhi Wang, Tanveer Syeda-Mahmood}
\institute{IBM Research -- Almaden Research Center, San Jose, CA, USA\\
\email{\{clwong, hongzhiw, stf\}@us.ibm.com}
}
\maketitle              
\begin{abstract}
With the introduction of Transformers, different attention-based models have been proposed for image segmentation with promising results. Although self-attention allows capturing of long-range dependencies, it suffers from a quadratic complexity in the image size especially in 3D. To avoid the out-of-memory error during training, input size reduction is usually required for 3D segmentation, but the accuracy can be suboptimal when the trained models are applied on the original image size. To address this limitation, inspired by the Fourier neural operator (FNO), we introduce the HartleyMHA model which is robust to training image resolution with efficient self-attention. FNO is a deep learning framework for learning mappings between functions in partial differential equations, which has the appealing properties of zero-shot super-resolution and global receptive field. We modify the FNO by using the Hartley transform with shared parameters to reduce the model size by orders of magnitude, and this allows us to further apply self-attention in the frequency domain for more expressive high-order feature combination with improved efficiency. When tested on the BraTS'19 dataset, it achieved superior robustness to training image resolution than other tested models with less than 1\% of their model parameters\footnote{GitHub repository: \url{https://github.com/IBM/multimodal-3d-image-segmentation}}.

\end{abstract}

\section{Introduction}

Convolutional neural networks (CNNs) have been widely used for medical image segmentation because of their speed and accuracy \cite{Journal:Hesamian:DI2019:deep,Journal:Liu:Sustainability2021:review}. Nevertheless, given the local receptive fields of convolutional layers, long-range spatial correlations are mainly captured through consecutive convolutions and pooling. For computationally demanding 3D segmentation, the receptive fields and abstract levels can be more limited than in 2D as fewer layers can be used. To balance between computational complexity and network capability, input size reductions by image downsampling and patch-wise training are common approaches. However, CNNs trained with downsampled images can be suboptimal when applied on the original resolution, and the receptive field of patch-wise training can be largely reduced depending on the patch size.

With the introduction of Transformers \cite{Conference:Vaswani:NIPS2017:attention} and their vision alternatives \cite{Conference:Dosovitskiy:ICLR2021:an,Conference:Liu:ICCV2021:swin}, the self-attention mechanism for long-range dependencies has been adopted to medical image segmentation with promising results \cite{Conference:Gao:MICCAI2021:utnet,Conference:Xie:MICCAI2021:cotr,Conference:Hatamizadeh:WACV2022:unetr,Workshop:Cao:ECCV2023:swin}. These approaches form a sequence of samples by either using the pixel values of low-resolution features \cite{Conference:Gao:MICCAI2021:utnet,Conference:Xie:MICCAI2021:cotr} or by dividing an image into smaller patches \cite{Conference:Hatamizadeh:WACV2022:unetr,Workshop:Cao:ECCV2023:swin}, and the multi-head attention is used to learn the dependencies among samples. Although these self-attention approaches allow capturing of long-range dependencies, as the computational requirements are proportional to sequence lengths and patch sizes which are proportional to image sizes, size-reduction approaches are needed for large images especially in 3D.

As image size reduction is usually required for large images, it is desirable to have a model that is robust to training image resolution so that the trained model can be applied to higher-resolution images with decent accuracy. Furthermore, as self-attention of Transformers allows better expressiveness through high-order channel and sample mixing \cite{Conference:Tolstikhin:NIPS2021:advances,Journal:Lee:Arxiv2021:fnet}, incorporating self-attention in an efficient way can be beneficial. To gain these advantages, here we propose the HartleyMHA model which is a resolution-robust and parameter-efficient network architecture with frequency-domain self-attention for 3D image segmentation. This model is based on the Fourier neural operator (FNO) \cite{Conference:Li:ICLR2021:fourier}, which is a deep learning model that learns mappings between functions in partial differential equations (PDEs) and has the appealing properties of zero-shot super-resolution and global receptive field. Our contributions include:
\begin{enumerate}
  \item To utilize the FNO for computationally expensive 3D segmentation, we modify it by using the Hartley transform with shared model parameters in the frequency domain. Residual connections \cite{Conference:He:CVPR2016} and deep supervision \cite{Conference:Lee:AISTATS2015} are also introduced. These reduce the number of model parameters by orders of magnitude and improve accuracy. We call it the HNOSeg model.
  \item As only low-frequency components are required for decent segmentation results, multi-head self-attention can be efficiently applied in the frequency domain. This allows high-order combination of features to improve the expressiveness of the model. We call it the HartleyMHA model.
  \item We compare our proposed models with other models on different training image resolutions to study their robustness. This provides useful insights that are usually unavailable in other studies.
\end{enumerate}
Experimental results on the BraTS'19 dataset \cite{Journal:Menze:TMI2015:multimodal,Journal:Bakas:SD2017:advancing,Journal:Bakas:arXiv2018:identifying} show that the proposed models have superior robustness to training image resolution than other tested models with less than 1\% of their model parameters.

\section{Methodology}

\subsection{Fourier Neural Operator}

FNO is a deep learning model for learning mappings between functions in PDEs without the PDEs provided \cite{Conference:Li:ICLR2021:fourier}. By formulating the solution in the continuous space based on the Green's function \cite{Journal:Li:arXiv2020:neural}, FNO can learn a single set of model parameters for multiple resolutions. For computationally expensive 3D segmentation, such zero-shot super-resolution capability is advantageous as a model trained with lower-resolution images can be applied on higher-resolution images with decent accuracy. The neural operator is formulated as iterative updates:
\begin{equation}\label{eq:fno_update}
    \begin{gathered}
      u_{t+1}(x) \coloneqq \sigma \left(W u_t(x) + \left(\mathcal{K}u_t\right)(x)\right) \\
      \textrm{with} \ \ \left(\mathcal{K}u_t\right)(x) \coloneqq \int_{D} \kappa(x - y)u_t(y) \,dy, \ \ \forall x \in D
    \end{gathered}
\end{equation}
where $u_t(x) \in \mathbb{R}^{d_{u_t}}$ is a function of $x$. $W \in \mathbb{R}^{d_{u_{t+1}} \times d_{u_t}}$ is a learnable linear transformation and $\sigma$ accounts for normalization and activation. In our work, $D \subset \mathbb{R}^3$ represents the 3D imaging space, and $u_t(x)$ are the outputs of hidden layers with $d_{u_t}$ channels. $\mathcal{K}$ is the kernel integral operator with $\kappa \in \mathbb{R}^{d_{u_{t+1}} \times d_{u_t}}$ a learnable kernel function. As $\left(\mathcal{K}u_t\right)$ is a convolution, it can be efficiently solved by the convolution theorem which states that the Fourier transform ($\mathcal{F}$) of a convolution of two functions is the pointwise product of their Fourier transforms:
\begin{equation}\label{eq:fourier_conv}
    \begin{split}
      \left(\mathcal{K}u_t\right)(x) = \mathcal{F}^{-1}\left(\mathcal{F}(\kappa) \mathcal{F}(u_t)\right)(x) = \mathcal{F}^{-1}\left(R U_t\right)(x), \ \ \forall x \in D
    \end{split}
\end{equation}
$R(k) = (\mathcal{F}\kappa)(k) \in \mathbb{C}^{d_{u_{t+1}} \times d_{u_t}}$ is a learnable function in the frequency domain and $U_t(k) = \left(\mathcal{F}u_t\right)(k) \in \mathbb{C}^{d_{u_t}}$. Therefore, each pointwise product at $k$ is realized as a matrix multiplication. When the fast Fourier transform is used in implementation, $k \in \mathbb{N}^3$ are non-negative integer coordinates, and each $k$ has a learnable $R(k)$. As mainly low-frequency components are required for image segmentation, only $k_i \leq k_{\mathrm{max},i}$ corresponding to the lower frequencies in each dimension $i$ are used to reduce model parameters and computation time.

\subsection{Hartley Neural Operator (HNO)}

As the FNO requires complex number operations in the frequency domain, the computational requirements such as memory and floating point operations are higher than with real numbers. Therefore, we use the Hartley transform instead, which is an integral transform alternative to the Fourier transform \cite{Journal:Hartley:IRE1942:more}. The Hartley transform ($\mathcal{H}$) converts real-valued functions to real-valued functions, which is related to the Fourier transform as $(\mathcal{H}f) = \mathrm{Real}(\mathcal{F}f) - \mathrm{Imag}(\mathcal{F}f)$. The convolution theorem of discrete Hartley transform is more complicated \cite{Journal:Bracewell:JOSA1983:discrete}, and the kernel integration in (\ref{eq:fno_update}) becomes:
\begin{equation}\label{eq:hartley_conv}
    \begin{gathered}
      \mathcal{H}\left(\mathcal{K}u_t\right)(k) = \frac{\hat{R}(k) \left(\hat{U}_t(k) + \hat{U}_t(N - k)\right) + \hat{R}(N - k) \left(\hat{U}_t(k) - \hat{U}_t(N - k)\right)}{2}
    \end{gathered}
\end{equation}
with $\hat{R}(k) = (\mathcal{H}\kappa)(k) \in \mathbb{R}^{d_{u_{t+1}} \times d_{u_t}}$ and $\hat{U}_t(k) = \left(\mathcal{H}u_t\right)(k) \in \mathbb{R}^{d_{u_t}}$. $N \in \mathbb{N}^3$ is the size of the frequency domain. $\hat{R}$ and $\hat{U}$ are $N$-periodic in each dimension \footnote{In Python, this means $\hat{U}[N_x, :, :] = \hat{U}[0, : , :]$, etc.}.

Similar to using (\ref{eq:fourier_conv}), the models built using (\ref{eq:hartley_conv}) have tens of million parameters even with small $k_\mathrm{max}$ (e.g., (14, 14, 10)). Therefore, instead of using a different $\hat{R}(k)$ at each $k$, we use the same (shared) $\hat{R}$ for all $k$ and (\ref{eq:hartley_conv}) becomes:
\begin{equation}\label{eq:hartley_conv_shared}
    \begin{gathered}
      \mathcal{H}\left(\mathcal{K}u_t\right)(k) = \hat{R} \hat{U}_t(k)
    \end{gathered}
\end{equation}
This is equivalent to applying a convolution layer with the kernel size of one in the frequency domain. We find that using (\ref{eq:hartley_conv_shared}) simplifies the computation and largely reduces the number of parameters without affecting the accuracy.

\subsection{Hartley Multi-Head Attention (MHA)}

As real instead of complex numbers are used in (\ref{eq:hartley_conv_shared}), multi-head attention in \cite{Conference:Vaswani:NIPS2017:attention} can be applied in the frequency domain for high-order feature combination. As $k_\mathrm{max}$ can be much smaller than the image size for image segmentation, the sequence length (number of voxels) can be largely reduced. With (\ref{eq:hartley_conv_shared}), the query, key, and value matrices ($Q$, $K$, $V$) of self-attention can be computed as:
\begin{equation}\label{eq:QKV}
    \begin{gathered}
      Q = \bar{U}_t \hat{R}_Q^\mathrm{T}, \ K = \bar{U}_t \hat{R}_K^\mathrm{T}, \ V = \bar{U}_t \hat{R}_V^\mathrm{T} \ \in \mathbb{R}^{N_f \times d_{u_{t+1}}}
    \end{gathered}
\end{equation}
where $\bar{U}_t \in \mathbb{R}^{N_f \times d_{u_t}}$, with $N_f$ = $8 k_{\mathrm{max},x} k_{\mathrm{max},y} k_{\mathrm{max},z}$, is a 2D matrix formed by stacking $\hat{U}_t(k)$ \footnote{Note that $k_\mathrm{max} = (k_{\mathrm{max},x}, k_{\mathrm{max},y}, k_{\mathrm{max},z})$ corresponds to a frequency domain of size $2k_{\mathrm{max},x} \times 2k_{\mathrm{max},y} \times 2k_{\mathrm{max},z}$ to cover both positive and negative frequency terms.}. Although $N_f$ can be relatively small, the computation and memory requirements of computing $QK^\mathrm{T}$ can still be demanding. For example, $k_\mathrm{max}$ = (14, 14, 10) corresponds to an attention matrix with around 246M elements. To remedy this, for each $Q$, $K$, and $V$, we group the feature vectors with a patch size of 2$\times$2$\times$2 voxels in the frequency domain and their matrix sizes become $\frac{N_f}{8} \times 8d_{u_{t+1}}$. This reduces the number of elements in $QK^\mathrm{T}$ by 64 times. The self-attention can then be computed as:
\begin{equation}\label{eq:attn}
    \begin{gathered}
      \textrm{Attention}(Q, K, V) = \textrm{SELU}\left(QK^\mathrm{T} / \sqrt{8d_{u_{t+1}}}\right) V \in \mathbb{R}^{\frac{N_f}{8} \times 8d_{u_{t+1}}}
    \end{gathered}
\end{equation}
where SELU represents the scaled exponential linear unit \cite{Conference:Klambauer:NIPS2017:self}. Similar to \cite{Conference:Lu:NIPS2021:soft}, we find that using softmax in self-attention results in suboptimal segmentations, thus the SELU was chosen after testing with multiple activations. Furthermore, we find that position encoding is unnecessary. The result of (\ref{eq:attn}) can be rearranged back to the original shape in the frequency domain so that the inverse Hartley transform can be applied. The multi-head attention can be used with (\ref{eq:attn}).

\begin{figure}[t]
    \centering
    \begin{minipage}[t]{1\linewidth}
      \includegraphics[width=\linewidth]{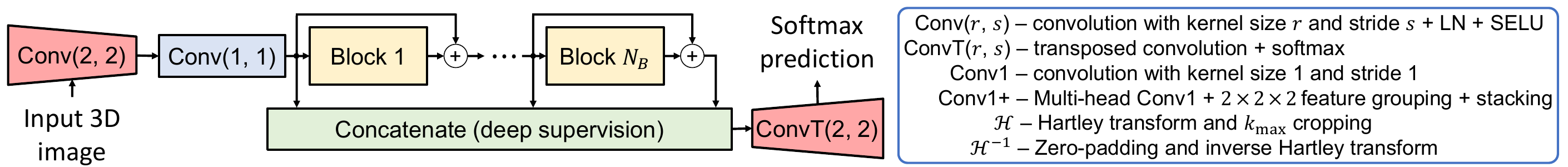}
    \end{minipage}
    \begin{minipage}[t]{1\linewidth}
      \includegraphics[width=0.46\linewidth]{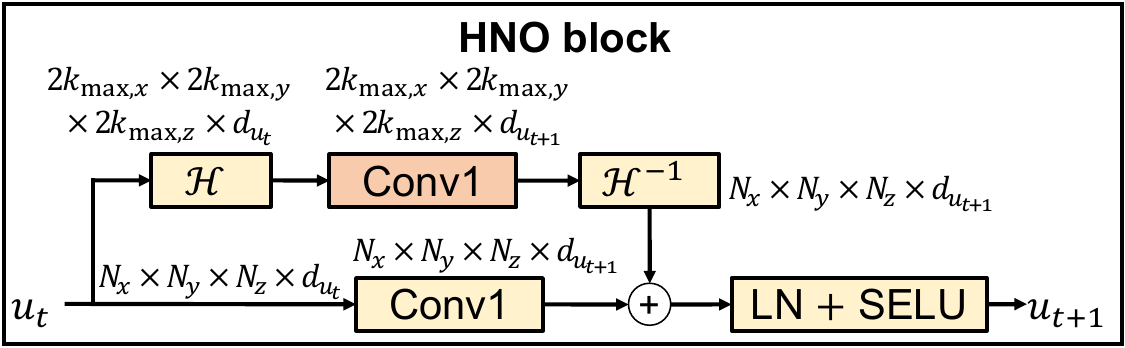}
      \hfill
      \includegraphics[width=0.52\linewidth]{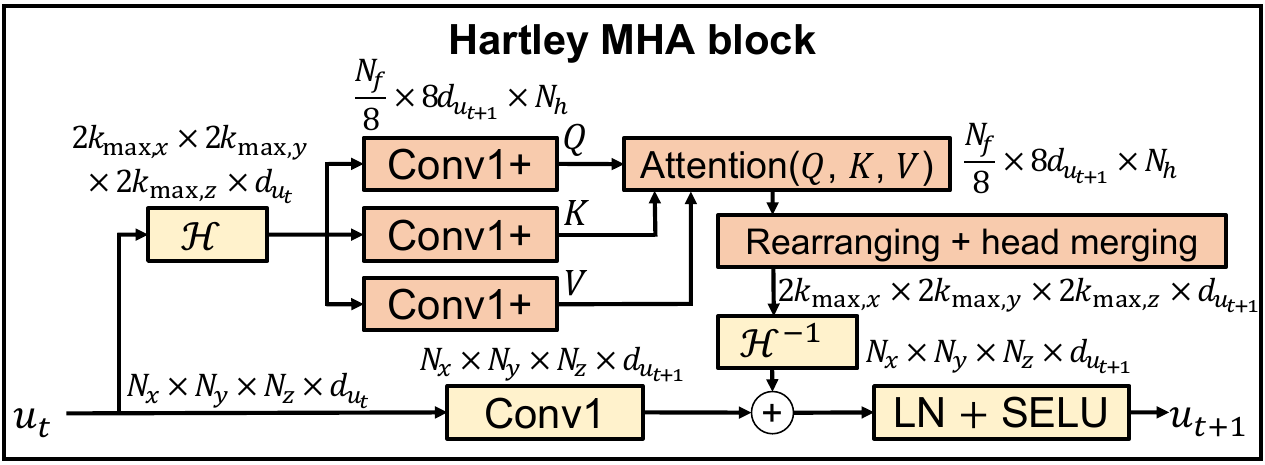}
    \end{minipage}
    \caption{Network architecture. The blocks are (\ref{eq:fno_update}) with the kernel integral operator implemented by the Hartley transform (HNO block) or the Hartley multi-head attention (Hartley MHA block). $N_h = 4$ is the number of heads. We use $d_{u_{t+1}}$ = $d_{u_{t}}$ = 12, $k_\mathrm{max}$ = (14, 14, 10), and $N_B$ = 32 with the HNO block and $N_B$ = 16 with the Hartley MHA block. The red blocks are for learnable resampling.}
    \label{fig:network}
\end{figure}

\subsection{Network Architectures -- HNOSeg and HartleyMHA}
\label{sec:network}

Fig. \ref{fig:network} shows the network architecture. We call it HNOSeg with the HNO blocks and HartleyMHA with the Hartley MHA blocks. Different from the FNO in \cite{Conference:Li:ICLR2021:fourier}, residual connections \cite{Conference:He:CVPR2016} and deep supervision \cite{Conference:Lee:AISTATS2015} are used to improve the training stability, convergence, and accuracy. As the batch size is usually small for memory demanding 3D segmentation, layer normalization (LN) is used \cite{Journal:Ba:arXiv2016:layer}. The SELU \cite{Conference:Klambauer:NIPS2017:self} is used as the activation function, and the softmax function is used to produce the final prediction scores. Similar to the Fourier transform, the Hartley transform provides a global receptive field as all voxels are used to compute the value at each $k$, thus pooling is not required. As using the original image resolution usually results in out-of-memory errors in 3D segmentation, downsampling the inputs and then upsampling the predictions may be required. Instead of using traditional image resampling methods, we use a convolutional layer with the kernel size and stride of two right after the input layer, and replace the output convolutional layer by a transposed convolutional layer with the kernel size and stride of two (red blocks in Fig. \ref{fig:network}). In this way, the model can learn the optimal resampling approach. In the experiments, $k_\mathrm{max}$ = (14, 14, 10) so that it can be used with the lowest tested training resolution of 60$\times$60$\times$39. Other hyperparameters such as $d_{u_{t+1}}$, $N_h$, and $N_B$  were obtained empirically for decent segmentations when training with the original image resolution. As each HNO block and Hartley MHA block can be implemented as a deep-learning layer in commonly used libraries, they can be easily adopted by other architectures.

\subsection{Training Strategy}

The images of different modalities are stacked along the channel axis to provide a multi-channel input. As the intensity ranges across modalities can be quite different, intensity normalization is performed on each image of each modality. Image augmentation with rotation (axial, $\pm$\ang{30}), shifting ($\pm$20\%), and scaling ([0.8, 1.2]) is used and each image has an 80\% chance to be transformed. The Adamax optimizer \cite{Journal:Kingma:arXiv2014} is used with the cosine annealing learning rate scheduler \cite{Conference:Loshchilov:ICLR2017:SGDR}, with the maximum and minimum learning rates as $10^{-2}$ and $10^{-3}$, respectively. The Pearson's correlation coefficient loss is used as it is robust to learning rate for image segmentation \cite{Workshop:Wong:MLMI2022:3d}, and it consistently outperformed the Dice loss and weighted cross-entropy in our experiments. An NVIDIA Tesla P100 GPU with 16 GB memory is used with a batch size of one and 100 epochs, and Keras in TensorFlow 2.6.2 is used for implementation. Note that small batch sizes are common in 3D segmentation given the high memory requirement.

\section{Experiments}
\label{sec:experiments}

\begin{figure}[t]
    \centering
    \begin{minipage}[h]{0.49\linewidth}
      \includegraphics[width=\linewidth]{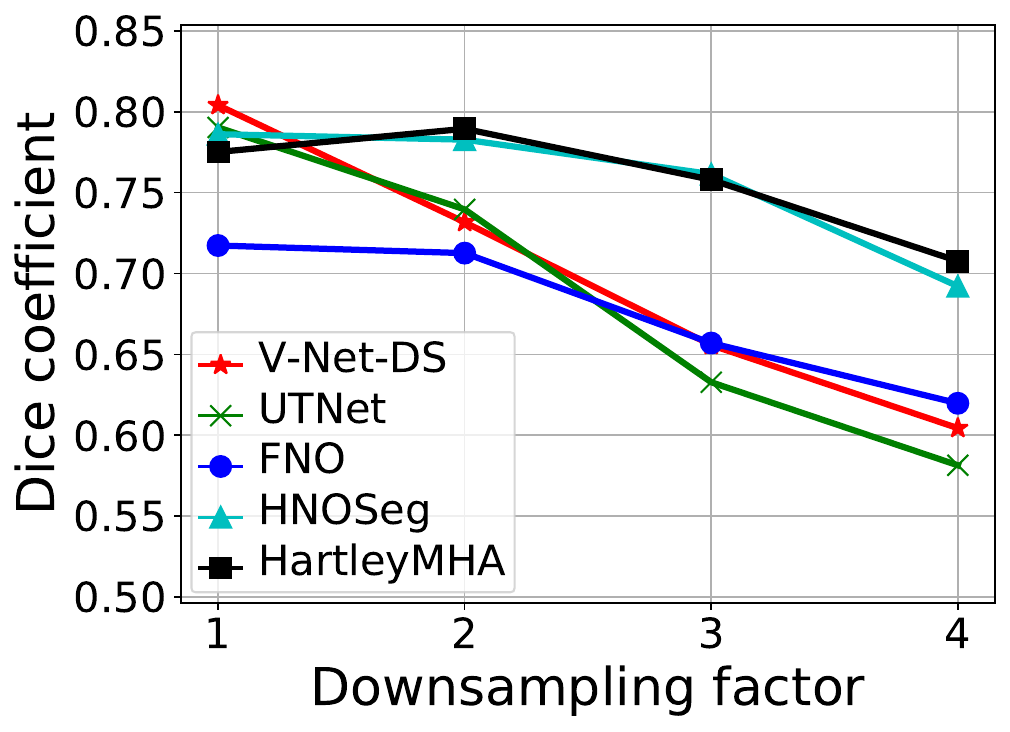}
    \end{minipage}
    \hfill
    \begin{minipage}[h]{0.47\linewidth}
      \includegraphics[width=\linewidth]{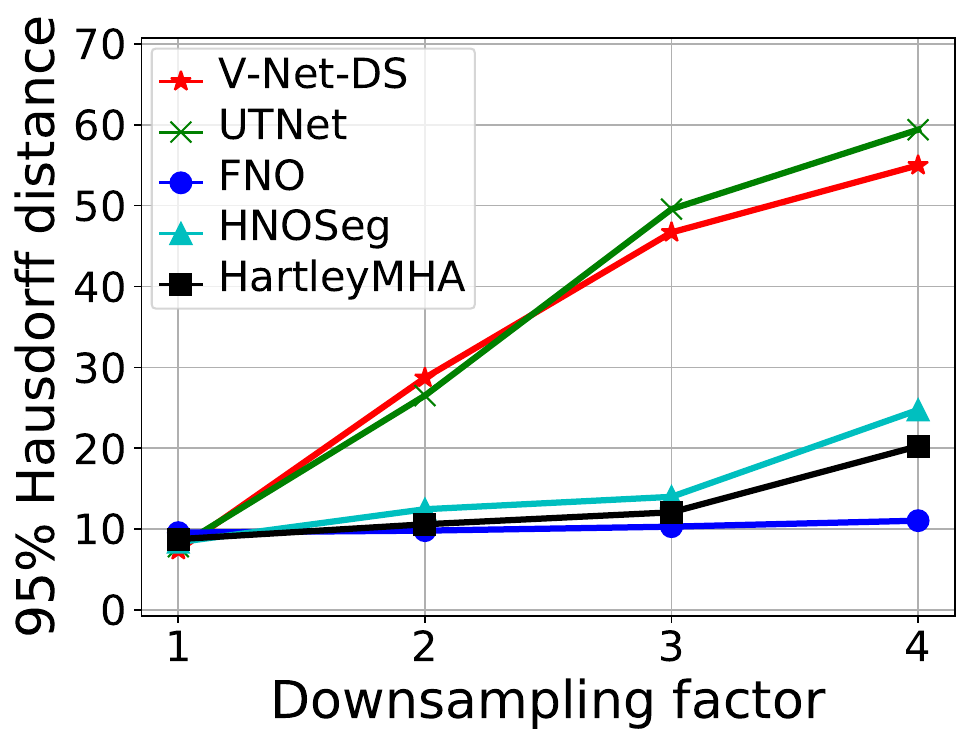}
    \end{minipage}
    \caption{Comparisons of robustness to training image resolution. Each point represents the average value from WT, TC, and ET of the 125 official validation cases of BraTS'19. The training images were downsampled by different factors while the trained models were tested with the original resolution.}
    \label{fig:plot_vs_downfactor}
\end{figure}

\subsection{Data and Experimental Setups}

The dataset of BraTS'19 with 335 cases of gliomas was used, each with four modalities of T1, post-contrast T1, T2, and T2-FLAIR images with 240$\times$240$\times$155 voxels \cite{Journal:Bakas:arXiv2018:identifying}. There is also an official validation dataset of 125 cases in the same format without given annotations. Models were trained with images downsampled by different factors (1, 2, 3, and 4) to study the robustness to image resolution. In training, we split the training dataset (335 cases) into 90\% for training and 10\% for validation. In testing, each model was tested on the official validation dataset (125 cases) with 240$\times$240$\times$155 voxels regardless of the downsampling factor. The predictions were uploaded to the CBICA Image Processing Portal\footnote{https://ipp.cbica.upenn.edu/} for the results statistics of the ``whole tumor'' (WT), ``tumor core'' (TC), and ``enhancing tumor'' (ET) regions \cite{Journal:Bakas:arXiv2018:identifying}. We compare our proposed HNOSeg and HartleyMHA models with three other models:
\begin{enumerate}
  \item \textbf{V-Net-DS} \cite{Conference:Wong:MICCAI2018}: a V-Net with deep supervision representing the commonly-used encoding-decoding architectures.
  \item \textbf{UTNet} \cite{Conference:Gao:MICCAI2021:utnet}: a U-Net enhanced by the Transformer's attention mechanism.
  \item \textbf{FNO} \cite{Conference:Li:ICLR2021:fourier}: original FNO without shared parameters, residual connections, and deep supervision. The same hyperparameters as HNOSeg were used.
\end{enumerate}
The learnable resampling approach in Section \ref{sec:network} was applied to all models. Note that our goal is not competing for the best accuracy but studying the robustness to image resolution. Although only the results of a dataset are shown because of the page limit, the characteristics of the proposed models can be demonstrated through this challenging multi-modal brain tumor segmentation problem.

\begin{table}[t]
\caption{Numerical comparisons of Dice coefficients (\%) and 95\% Hausdorff distances (HD95) with different training image resolutions. }
\label{table:results}

\fontsize{7}{8}\selectfont
\centering

\newcolumntype{C}{>{\raggedleft\arraybackslash}X@{\extracolsep{3pt}}}
\newcolumntype{P}{>{\raggedleft\arraybackslash}X@{\extracolsep{6pt}}}
\newcolumntype{L}{l@{\extracolsep{3pt}}}
\newcommand{\boldblue}[1]{\textcolor{blue}{\textbf{#1}}}

\begin{tabularx}{\linewidth}{LCCPCCPCCPCCC}
\toprule
Downsampling factor & \multicolumn{6}{c}{1 (240$\times$240$\times$155)} & \multicolumn{6}{c}{3 (80$\times$80$\times$52)} \\
\cline{1-1} \cline{2-7} \cline{8-13} \noalign{\smallskip}
Metric & \multicolumn{3}{c}{Dice (\%)} & \multicolumn{3}{c}{HD95} & \multicolumn{3}{c}{Dice (\%)} & \multicolumn{3}{c}{HD95} \\
\cline{1-1} \cline{2-4} \cline{5-7} \cline{8-10} \cline{11-13} \noalign{\smallskip}
Region & WT & TC & ET & WT & TC & ET & WT & TC & ET & WT & TC & ET \\
\midrule
V-Net-DS
& 88.8 & 77.7 & 74.7 & 7.1 & 8.9 & 6.2
& 70.6 & 57.9 & 68.2 & 51.9 & 53.1 & 35.2
\\
UTNet
& 86.9 & 76.1 & 74.0 & 7.5 & 9.4 & 6.6
& 69.9 & 56.8 & 63.1 & 49.4 & 57.0 & 42.3
\\
FNO
& 84.0 & 69.0 & 62.2 & 9.4 & 11.2 & 8.1
& 79.3 & 63.8 & 54.0 & 10.0 & 11.2 & 9.8
\\
HNOSeg
& 87.7 & 75.0 & 73.2 & 9.1 & 9.6 & 6.5
& 86.7 & 71.9 & 69.8 & 15.0 & 14.4 & 12.6
\\
HartleyMHA
& 86.9 & 73.1 & 72.5 & 9.2 & 9.8 & 7.3
& 84.8 & 72.8 & 69.8 & 12.5 & 13.1 & 10.7
\\
\bottomrule
\end{tabularx}
\end{table}

\begin{table}[t]
\caption{Number of parameters, inference time per image in seconds averaged from images of size 240$\times$240$\times$155, and memory in GB with a batch size of 1.}
\label{table:inference}

\fontsize{7}{8}\selectfont
\centering

\newcolumntype{C}{>{\centering\arraybackslash}X@{\extracolsep{3pt}}}
\newcolumntype{P}{>{\centering\arraybackslash}X@{\extracolsep{8pt}}}
\newcommand{\boldblue}[1]{\textcolor{blue}{\textbf{#1}}}

\begin{tabularx}{\linewidth}{PCPPCPPCPPCPPCC}
\toprule
\multicolumn{3}{c}{V-Net-DS} & \multicolumn{3}{c}{UTNet} & \multicolumn{3}{c}{FNO} & \multicolumn{3}{c}{HNOSeg} & \multicolumn{3}{c}{HartleyMHA} \\
\cline{1-3} \cline{4-6} \cline{7-9} \cline{10-12} \cline{13-15} \noalign{\smallskip}
Param & Time & Mem & Param & Time & Mem & Param & Time & Mem & Param & Time & Mem & Param & Time & Mem \\
\midrule
5.7M & 0.33 & 9.0 & 7.1M & 0.41 & 4.9 & 144.5M & 0.61 & 4.8 & 24.8k & 0.71 & 8.9 & 47.7k & 0.57 & 4.8\\
\bottomrule
\end{tabularx}
\end{table}

\subsection{Results and Discussion}

Fig. \ref{fig:plot_vs_downfactor} and Table \ref{table:results} show comparisons of resolution robustness among tested models, and Table \ref{table:inference} shows the computational costs during inference. At the original resolution, V-Net-DS and UTNet outperformed HNOSeg and HartleyMHA by less than 3\% in the Dice coefficient on average, but HNOSeg and HartleyMHA only had less than 50k model parameters which were less than 1\% of V-Net-DS and UTNet. FNO performed worst with the most parameters (144.5M). As the resolution decreased, the accuracies of V-Net-DS and UTNet decreased almost linearly with the downsampling factor, while HNOSeg and HartleyMHA were more robust. When the downsampling factor changed from 1 to 3, the average Dice coefficients of V-Net-DS and UTNet decreased by more than 14.8\%, while those of HNOSeg and HartleyMHA only decreased by less than 2.5\%. Similar trends can be observed for the 95\% Hausdorff distance, except that FNO performed surprisingly well in this aspect. HartleyMHA performed better overall than HNOSeg. Note that we fixed $k_\mathrm{max}$ for the consistency among models in the experiments, which can be adjusted for better results in other situations.

For computation cost, Table \ref{table:inference} shows that V-Net-DS and UTNet had shorter inference times than HNOSeg and HartleyMHA, though all models used less than 0.8 seconds per image of size 240$\times$240$\times$155. HartleyMHA ran faster than HNOSeg and used less memory, though HartleyMHA had more parameters.

\begin{figure}[t]
\fontsize{6}{7}\selectfont
    \centering
    \begin{minipage}[t]{0.1\linewidth}
      \centering{Ground truth}
      \includegraphics[width=\linewidth]{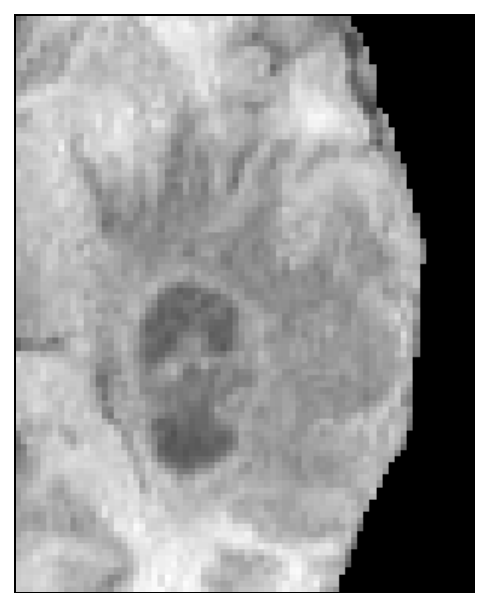} \\
      \includegraphics[width=\linewidth]{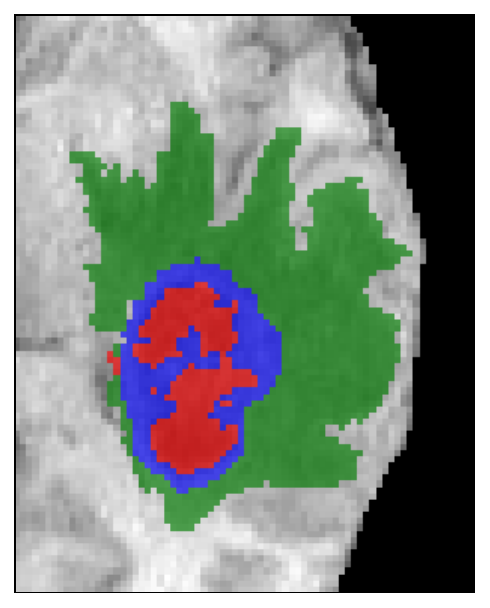}
    \end{minipage}
    \vline
    \vspace{1em}
    \begin{minipage}[t]{0.05\linewidth}
      \centering{\phantom{aaa}} \\
      \begin{minipage}[t][7em][t]{\linewidth}
        \rotatebox[origin=c]{90}{240$\times$240$\times$155}
      \end{minipage}
      \hfill \\
      \begin{minipage}[t][7em][t]{\linewidth}
        \rotatebox[origin=c]{90}{120$\times$120$\times$78}
      \end{minipage}
      \hfill \\
      \begin{minipage}[t][6em][t]{\linewidth}
        \rotatebox[origin=c]{90}{80$\times$80$\times$52}
      \end{minipage}
      \hfill \\
      \begin{minipage}[t][6em][t]{\linewidth}
        \rotatebox[origin=c]{90}{60$\times$60$\times$39}
      \end{minipage}
    \end{minipage}
    \hspace{-2em}
    \begin{minipage}[t]{0.1\linewidth}
      \centering{V-Net-DS} \\
      \includegraphics[width=\linewidth]{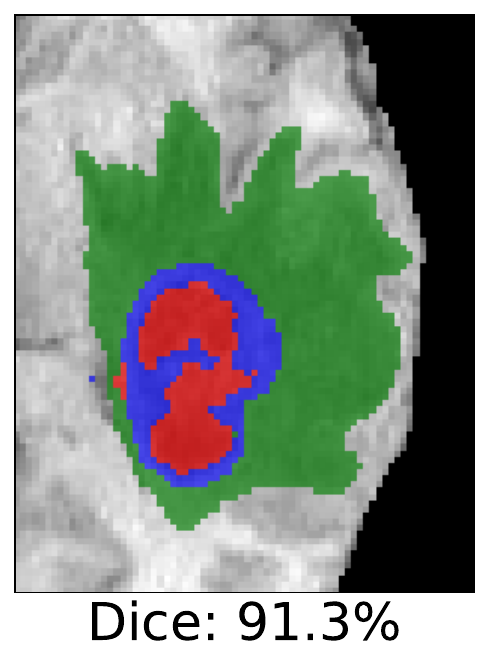} \\
      \includegraphics[width=\linewidth]{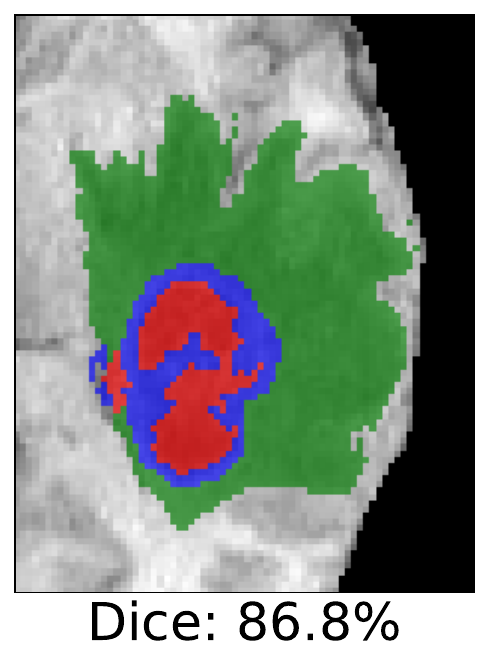} \\
      \includegraphics[width=\linewidth]{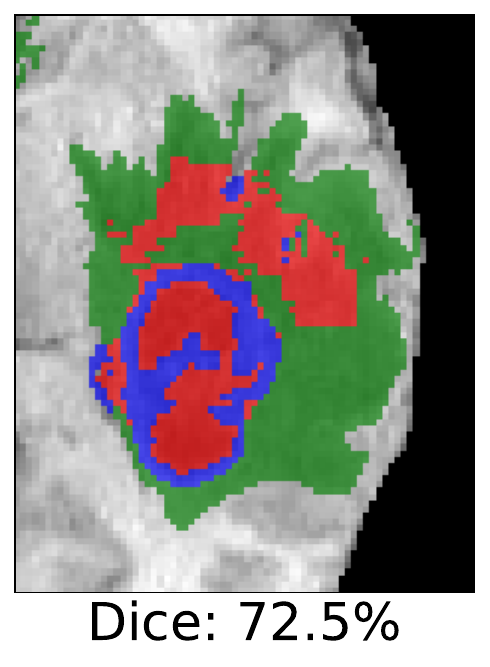} \\
      \includegraphics[width=\linewidth]{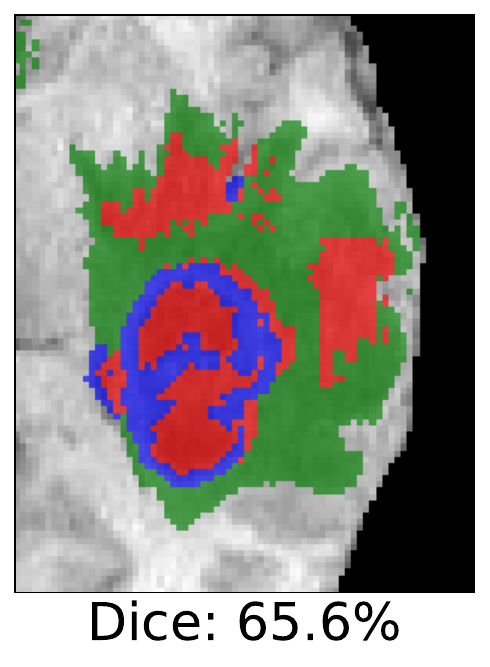}
    \end{minipage}
    \begin{minipage}[t]{0.1\linewidth}
      \centering{UTNet} \\
      \includegraphics[width=\linewidth]{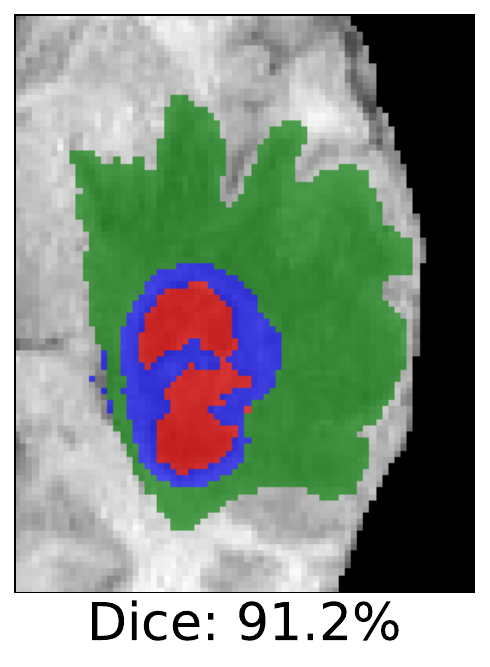} \\
      \includegraphics[width=\linewidth]{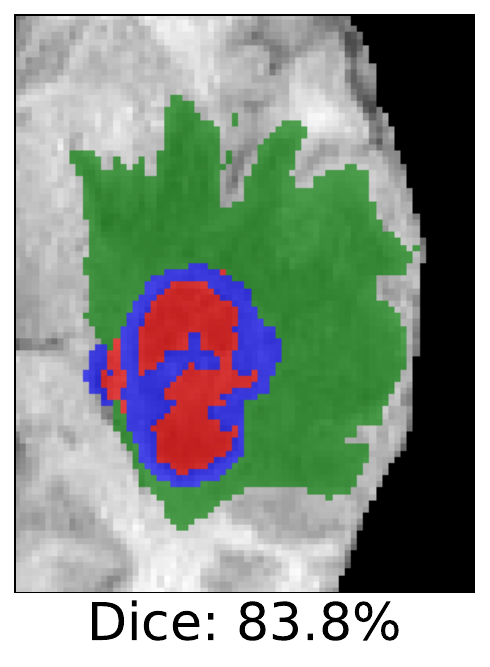} \\
      \includegraphics[width=\linewidth]{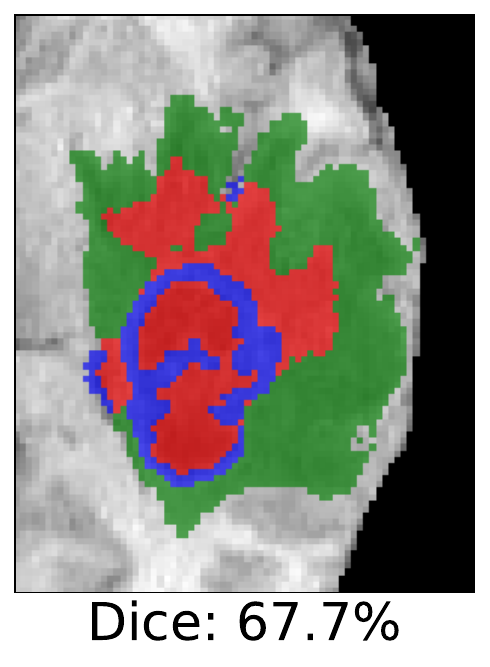} \\
      \includegraphics[width=\linewidth]{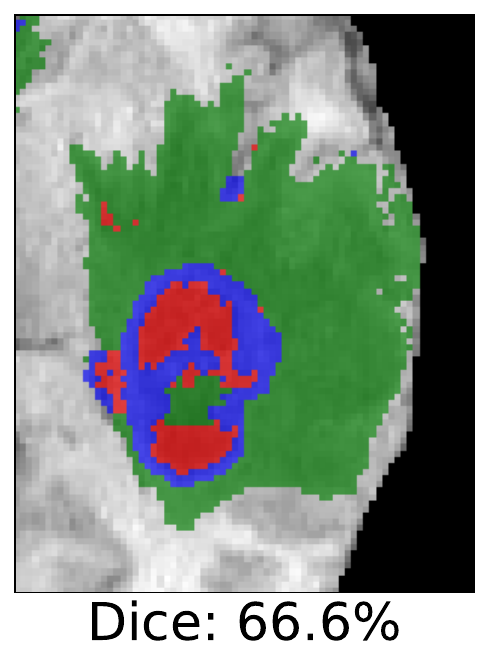}
    \end{minipage}
    \begin{minipage}[t]{0.1\linewidth}
      \centering{FNO} \\
      \includegraphics[width=\linewidth]{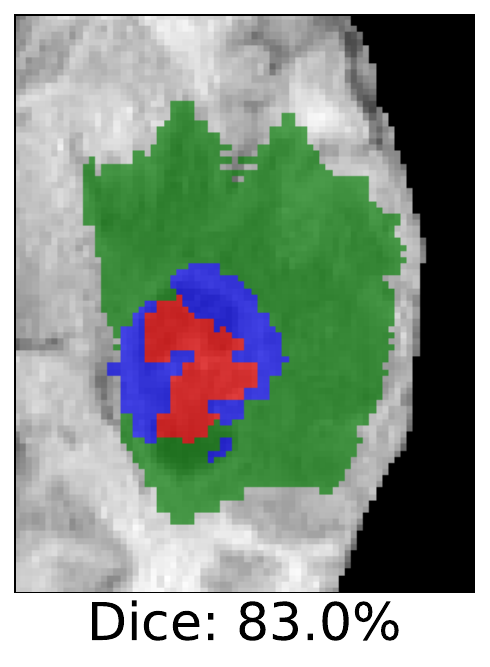} \\
      \includegraphics[width=\linewidth]{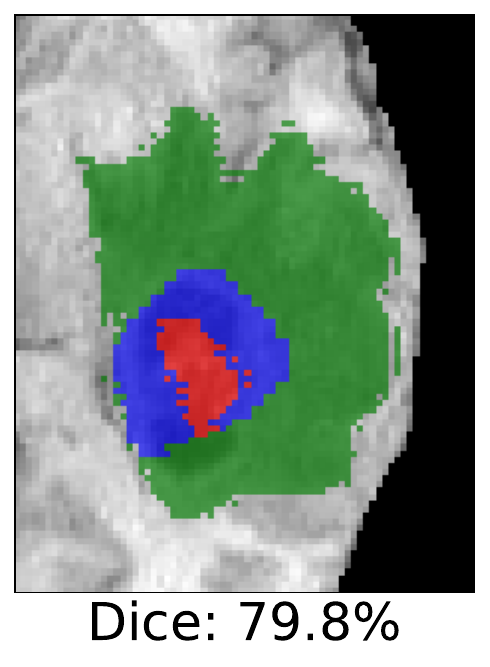} \\
      \includegraphics[width=\linewidth]{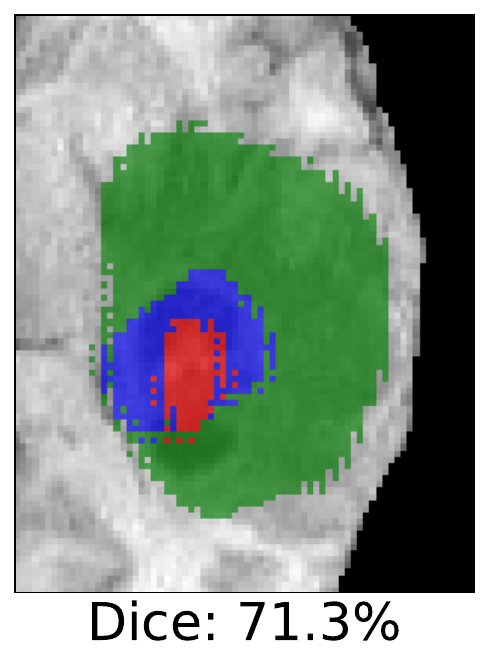} \\
      \includegraphics[width=\linewidth]{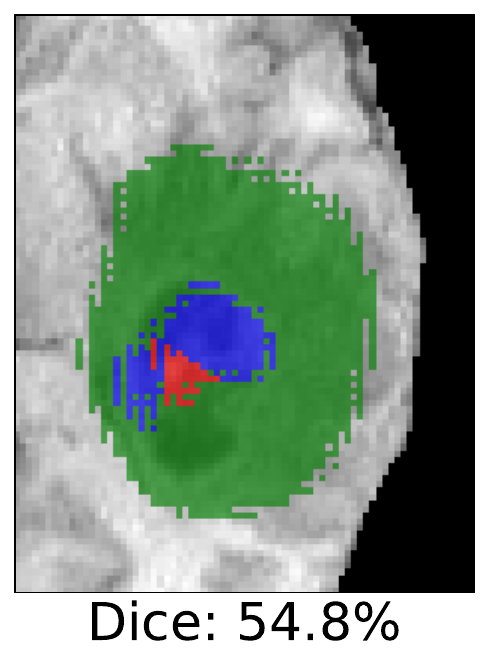}
    \end{minipage}
    \begin{minipage}[t]{0.1\linewidth}
      \centering{HNOSeg} \\
      \includegraphics[width=\linewidth]{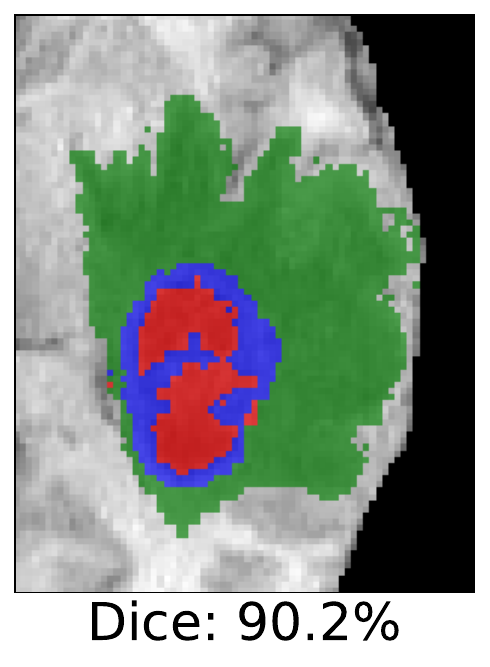} \\
      \includegraphics[width=\linewidth]{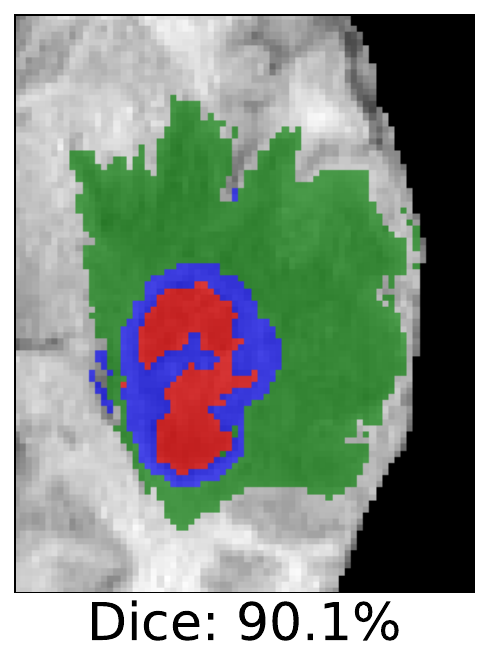} \\
      \includegraphics[width=\linewidth]{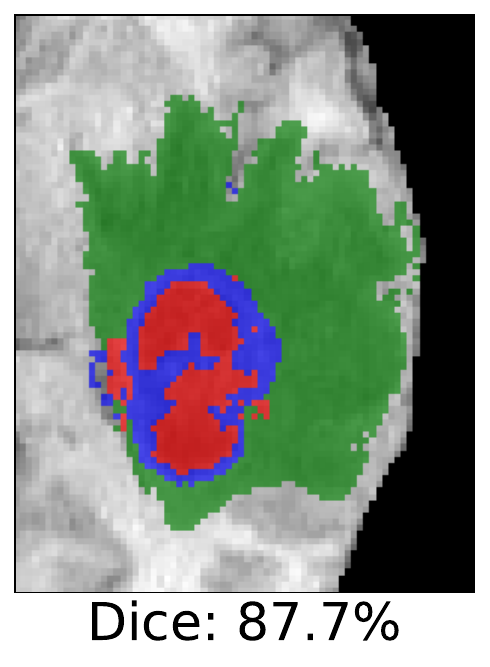} \\
      \includegraphics[width=\linewidth]{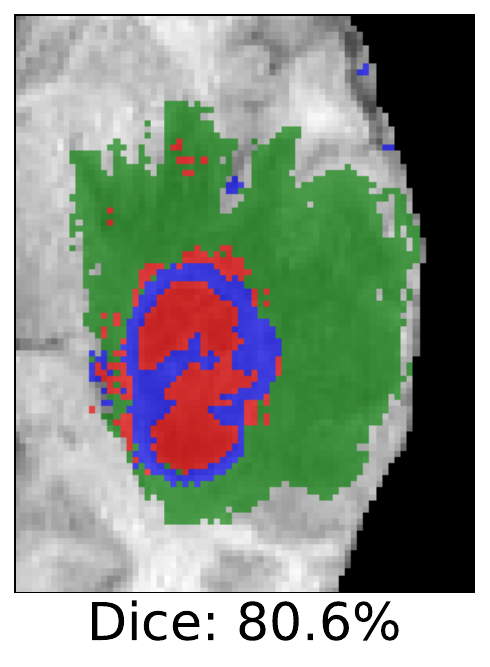}
    \end{minipage}
    \begin{minipage}[t]{0.1\linewidth}
      \centering{HartleyMHA} \\
      \includegraphics[width=\linewidth]{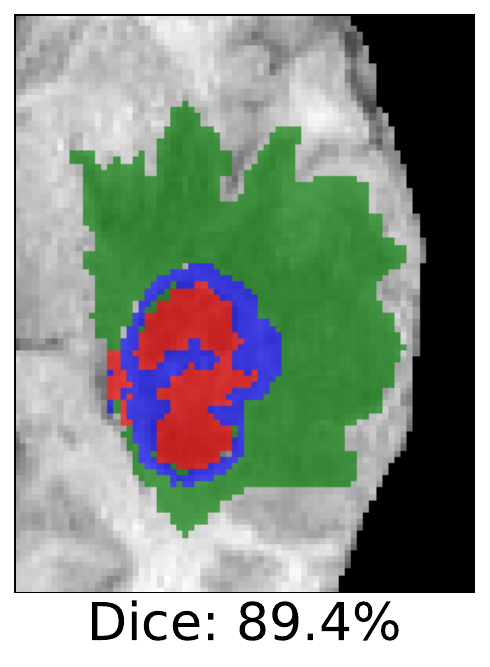} \\
      \includegraphics[width=\linewidth]{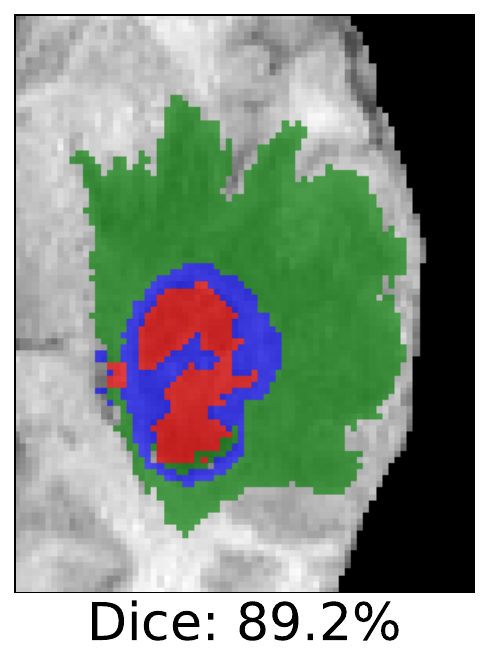} \\
      \includegraphics[width=\linewidth]{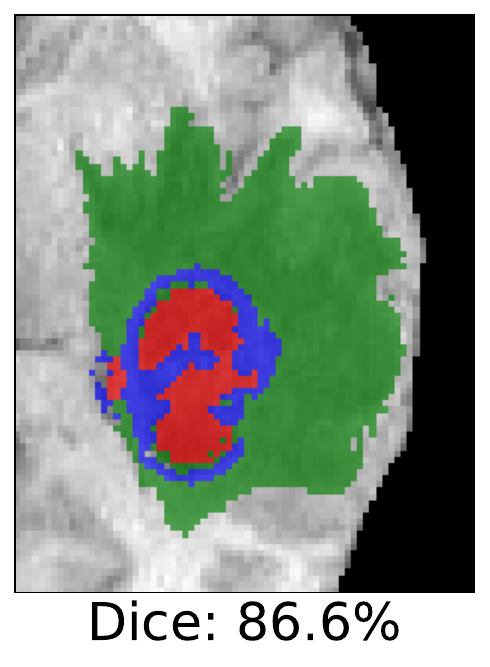} \\
      \includegraphics[width=\linewidth]{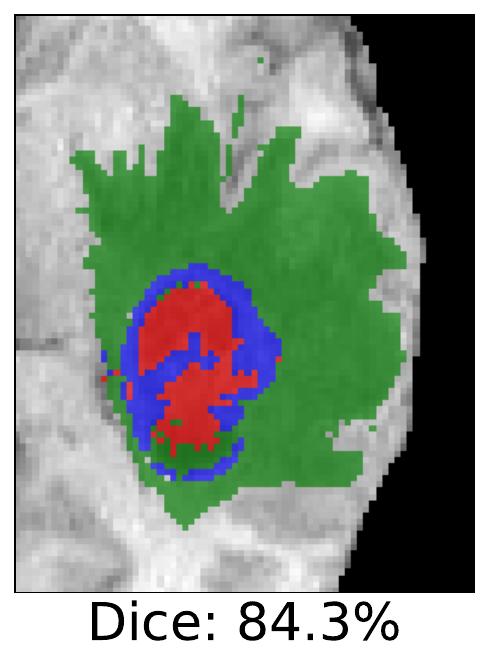}
    \end{minipage}
    \caption{Visual comparisons among models trained with different image resolutions, tested on an unseen case of size 240$\times$240$\times$155. The Dice coefficients were averaged from the WT, TC, and ET regions.}
    \label{fig:visualization}
\end{figure}

Fig. \ref{fig:visualization} shows the visual comparisons of the segmentation results on an unseen case. Consistent to Fig. \ref{fig:plot_vs_downfactor}, except FNO, the accuracies of the models were similar with the original training image resolution. As the training image resolution reduced, HNOSeg and HartleyMHA gradually outperformed V-Net-DS and UTNet. When the training images were downsampled from 240$\times$240$\times$155 to 60$\times$60$\times$39 (downsampling factor = 4), the average Dice coefficients of V-Net-DS and UTNet decreased by more than 24\%, and HartleyMHA had the least reduction of 5.1\%.

With such superior robustness to image resolution, HNOSeg and HartleyMHA can be trained with lower-resolution images using fewer computational resources to provide decent segmentation results on the original resolution during inference. While HartleyMHA performed better than HNOSeg in general, their similar performance is consistent with the findings in \cite{Conference:Tolstikhin:NIPS2021:advances,Journal:Lee:Arxiv2021:fnet} that self-attention is sufficient for good performance but is not crucial. On the other hand, as the use of efficient self-attention improves the expressiveness of the Hartley MHA block, fewer layers can be used to reduce the overall computational costs.

\section{Conclusion}

In this paper, based on the idea of FNO which has the properties of zero-shot super-resolution and global receptive field, we propose the HNOSeg and HartleyMHA models for resolution-robust and parameter-efficient 3D image segmentation. HNOSeg is FNO improved by the Hartley transform, residual connections, deep supervision, and shared parameters in the frequency domain. We further extend this concept for efficient multi-head attention in the frequency domain as HartleyMHA. Experimental results show that HNOSeg and HartleyMHA had similar accuracies as other tested segmentation models when trained with the original image resolution, but had superior performance when trained with images of much lower resolutions. HartleyMHA performed slightly better than HNOSeg and ran faster with less memory. With these advantages, HartleyMHA can be a promising alternative for 3D image segmentation especially when computational resources are limited.

\bibliographystyle{splncs04}
\bibliography{Ref}

\end{document}